\documentclass[runningheads]{llncs}
\usepackage[T1]{fontenc}
\usepackage{graphicx}
\usepackage{amsmath}
\usepackage{amssymb}
\usepackage[shortlabels]{enumitem}
\usepackage{hyperref}
\usepackage[nameinlink]{cleveref}
\usepackage{color}

\def\totp{\mathsf{TotP}}
\def\sp{\mathsf{\#P}}
\def\gapp{\mathsf{GapP}}

\def\spanl{\mathsf{SpanL}}
\def\np{\mathsf{NP}}

\def\up{\mathsf{UP}}

\def\spp{\mathsf{SPP}}
\def\pp{\mathsf{PP}}
\def\cp{\mathsf{C_{=}P}}

\def\spe{\mathsf{\#PE}}
\def\fp{\mathsf{FP}}
\def\fpl{\mathsf{FP_{+}}}
\def\p{\mathsf{P}}

\newtheorem{Theorem}{Theorem}[section]
\newtheorem{Lemma}[Theorem]{Lemma}
\newtheorem{Proposition}[Theorem]{Proposition}
\newtheorem{Definition}[Theorem]{Definition}
\newtheorem{Corollary}[Theorem]{Corollary}

\begin{document}
	\title{Closure Properties and Characterizations of TotP}

	%
	%
	\author{Yaroslav Ivanashev} 
	%
	\authorrunning{Y. Ivanashev}
	%
	
	\institute{HSE University, Faculty of Computer Science, Moscow, 101000, Russia
		\email{yivanashev@hse.ru}}
	%
	\maketitle              
	\begin{abstract}
		The class $\totp$ consists of functions that count the number of all paths of a nondeterministic polynomial-time Turing machine. In this paper, we give a predicate based definition of $\totp$, analogous to a standard definition of $\sp$. From a new characterization of $\totp$ it follows that many well known $\sp$ problems belong to $\totp$, and $\totp = \sp$ if and only if $\p = \np$. We show that $\totp$ has several closure properties of $\sp$ and $\gapp$, and also properties that are not known to hold for $\sp$ and $\gapp$. We also prove that the closure of $\totp$ under left composition with $\fp_{+}$ is equivalent to $\totp = \fp_{+}$ and $\p = \pp$, and give examples of $\fp_{+}$-functions such that if $\totp$ is closed under composition with them, then it is closed under composition with $\fp_{+}$.

		
	\end{abstract}
	\section{Introduction}
	
	The class $\totp$, which was introduced in \cite{kia01}, is a subclass of $\sp$ \cite{val79} containing functions that count the number of all paths of a nondeterministic polynomial-time Turing machine (NPTM). In \Cref{sec-cha}, we give a predicate based definition of $\totp$, in which it is described as a class of functions equal to the number of witnesses for $\np$-problems of a certain type. 	
	From a new characterization of $\totp$ it follows that many $\sp$-problems belong to $\totp$, including $\textsc{NonNegative Permanent}$, $\textsc{\#Perfect Matchings}$, $\textsc{\#DNF-Sat}$, and $\textsc{\#NonCliques}$, which are $\sp$-complete under 1-Cook reductions. It also follows that $\totp = \sp$ if and only if $\p = \np$. 
	$\totp$ has complete problems under Karp reductions \cite{ant22}, and it coincides with the closure under Karp reductions of the class of self-reducible functions from the class $\spe$ \cite{pag06}. $\totp$ is characterized using interval-size functions in \cite{bam16}.
	In \cite{bak20}, the relationship of $\totp$ with classes of approximable functions is studied. Logical characterizations of $\totp$ and some of its subclasses are given in \cite{ach23,are20,bak20}. 

	$\totp$ has natural closure properties, for example, it is closed under addition, multiplication, and decrement by one \cite{bak23}. In \Cref{sec-clo}, we give several closure properties of $\totp$ that $\sp$ and $\gapp$ have \cite{bei92,fen94}, and also properties that are unknown to hold for $\sp$ and $\gapp$. In \Cref{sec-gap}, we give a characterization of $\gapp$ with $\totp$-functions that is more precise than $\gapp = \totp - \fp$ \cite{bak23}.
	
	
	In \Cref{sec-comp}, we study the closure of $\totp$ under composition with $\fpl$ (i.e., whether $\fpl \circ \totp \subseteq \totp$), and prove that this statement is equivalent to $\totp = \fpl$ and $\p = \pp$. For the class $\sp$, the closure under left composition with $\fp_{+}$ is equivalent to $\pp = \up$ and $\sp = \mathsf{UPSV_{t}}$ \cite{ogi93,iva25}; for $\gapp$ is equivalent\footnote{For $\gapp$, the closure under decrement by one is also equivalent to $\pp = \spp$ \cite{thi94}. For the class $\sp$, it is only known that the closure under decrement by one implies $\np \subseteq \spp$ \cite{ogi93}.} to $\pp = \spp$ and $\gapp = \mathsf{\fp^{\spp}}$ \cite{gup91,fen94,thi94,iva25}.
	
	We also give examples of $\fpl$-functions of several arguments such that if $\totp$ is closed under composition with them, then it is closed under composition with $\fp_{+}$. 
	Our proof is similar to that given in \cite{ogi93}, where analogous result for the class $\sp$ is obtained.
	
	\section{Preliminaries}
	
	In this paper, we use standard definitions of a deterministic and a nondeterministic Turing machine. Below we give definitions of function and language classes that we will consider:
	
	\begin{Definition}
		\begin{enumerate}
			\item  \textnormal{\cite{val79}} $\sp = \{acc_{M} \ | \ M$ is a NPTM$\}$, where $acc_{M}(x)$ is the number of accepting paths of $M$ on input $x \in \Sigma^{*}$.		
			\item \textnormal{\cite{fen94}} $\gapp = \{gap_{M} \ | \ M$ is a NPTM$\}$, where $gap_{M}(x) = acc_{M}(x) - rej_{M}(x)$ and $rej_{M}(x)$ is the number of rejecting paths of $M$ on input $x \in \Sigma^{*}$.
			\item \textnormal{\cite{kia01}} $\totp = \{tot_{M} \ | \ M$ is a NPTM$\}$, where $tot_{M}(x) = Total_{M}(x) - 1$ and $Total_{M}(x)$ is the number of all paths of $M$ on input $x \in \Sigma^{*}$.\footnote{The "-1" in the definition of $\totp$ allows these functions to take a zero value.}		
			\item \textnormal{\cite{pag06}} $\spe = \{f \ | \ f \in \sp$ and $\{x \ | \ f(x) > 0\} \in \p\}$.	
			\item $\fp = \{f: \Sigma^{*} \rightarrow \mathbb{Z} \ | \ f$ is computable in polynomial time$\}$.		
			\item $\fpl = \{f \ | \ f \in \fp$ and $f \geqslant 0\}$.
		\end{enumerate}
	\end{Definition}
	
	
	\begin{Definition}
		\begin{enumerate}
			\item \textnormal{\cite{coo71,lev73}} The class $\np$ consists of all languages $L$ for which there exists a function $f \in \sp$ such that for any $x \in \Sigma^{*}$: 
			\begin{align*}
				x \in L \Leftrightarrow f(x) > 0.
			\end{align*}
			\item \textnormal{\cite{sim75,wag86}} The class $\cp$ consists of all languages $L$ for which there exist functions $f \in \sp$ and $g \in \fp$ such that for any $x \in \Sigma^{*}$:
			\begin{align*}
				x \in L \Leftrightarrow f(x) = g(x).
			\end{align*}
			\item \textnormal{\cite{sim75,gil77}} The class $\pp$ consists of all languages $L$ for which there exist functions $f \in \sp$ and $g \in \fp$ such that for any $x \in \Sigma^{*}$:
			\begin{align*}
				x \in L \Leftrightarrow f(x) > g(x).
			\end{align*}
		\end{enumerate}
	\end{Definition}
			
	The relations between classes of functions are given in \Cref{rel-fun}: 
	
	\begin{Proposition} \label[Proposition]{rel-fun}
		\begin{enumerate}
			\item \textnormal{\cite{pag06}} $\fp_{+} \subseteq \totp \subseteq \spe \subseteq \sp$. All inclusions are proper	unless $\p = \np$.			
			\item \textnormal{\cite{fen94,bak23}} $\gapp  = \sp - \sp = \sp - \fp = \totp - \totp = \totp - \fp$. 
		\end{enumerate}
	\end{Proposition}
		
	\section{Characterizations of TotP} \label{sec-cha}
	
	In this section, we give a predicate based definition of $\totp$, analogous to the following definition of $\sp$:	
	
	\begin{Proposition} 
		The class $\sp$ consists of all functions $f$ for which there exist a polynomial $p$ and a language $A \in \p$ such that for any $x \in \Sigma^{*}$: 
		\begin{align*}
			f(x) = |\{y \ | \ |y| = p(|x|) \wedge (x, y) \in A\}|.
		\end{align*}
	\end{Proposition} 
	
	To obtain a characterization of $\totp$, we add an additional condition to the above characterization of $\sp$: there exists an algorithm that on any input $(x, c) \in \Sigma^{*}$ determines in polynomial time the existence of a witness for $x$ with a given prefix~$c$.
	
	\begin{Theorem} \label{pred-def}
		The class $\totp$ consists of all functions $f$ for which there exist a polynomial $p$ and a language $A \in \p$ such that:
		\begin{enumerate}
			\item For any $x \in \Sigma^{*}$: $f(x) = |\{y \ | \ |y| = p(|x|) \wedge (x, y) \in A\}|$.
			\item The language
			$\{(x, c) \ | \ \exists z (|cz| = p(|x|) \wedge (x, cz) \in A)\} \in \p$, where $cz$ denotes a concatenation of strings $c$ and $z$. 
		\end{enumerate}
	\end{Theorem}
	
	\begin{proof}
		$(\subseteq)$: Let $f(x) = tot_{M}(x)$ for a NPTM $M$. There a exists a a polynomial-time algorithm that accepts $(x, y)$ if and only if $y$ is a path of $M(x)$ that is not the lexicographically largest path of $M(x)$, where paths of $M(x)$ are encoded so that their length is always equal to a fixed polynomial $p(|x|)$. A polynomial-time algorithm can determine whether a given string is a prefix of a path of $M(x)$ that is not the lexicographically largest path of $M(x)$.
		
		$(\supseteq)$: Let $f$ be a function such that the above conditions hold. Then $f(x) = tot_{M}(x)$, where $M$ is the NPTM that behaves as follows on input $x \in \Sigma^{*}$:
		\begin{enumerate}[(a)]
			\item If the empty string is not a prefix of any witness for $x$, $M$ halts.
			\item $M$ takes one nondeterministic step and halts on the first path.
			\item On the second path, $M$ nondeterministically guesses a witness for $x$. Initially, $y$ is the empty string. If y0 and y1 are both prefixes of a witness for $x$, $M$ takes one nondeterministic step. On one path, it replaces $y$ with $y1$, and on the other path, it replaces $y$ with $y0$. If $y$ can be extended only in one way, $M$ doesn't take a nondeterministic step.
			\item If the length of $y$ becomes equal to $p(|x|)$, $M$ halts. \qed 
		\end{enumerate}
	\end{proof}
	
	If $\p = \np$, then the second condition in \Cref{pred-def} holds for all $\sp$-functions. This gives the following corollary: 
	
	\begin{Corollary} 
		$\totp = \sp$ if and only if $\p = \np$. 
	\end{Corollary} 
	
	From \Cref{pred-def} it follows\footnote{In \cite[p.413]{val79b}, Valiant observes that many $\sp$-complete problems have the second property from \Cref{pred-def}. It follows from Fact 7 on p.413 that for all $\totp$-problems there exists an algorithm that for any $x \in \Sigma^{*}$ outputs all witnesses in time $p(|x|)\cdot N$, where $N$ is the number of witnesses and $p$ is some polynomial (a term "algorithm with a polynomial delay" is used in \cite{joh88}). For all $\sp$-problems the same is not true unless $\p = \up$ \cite{val79}.} that many well known $\sp$-problems belong to $\totp$, including $\textsc{\#Perfect Matchings}$, $\textsc{NonNegative Permanent}$, \\ $\textsc{\#DNF-Sat}$, $\textsc{\#MonotoneSat}$, $\textsc{\#NonCliques}$, $\textsc{\#NonIndependent Sets}$, \\ $\textsc{\#NFA}$, which are $\sp$-complete under 1-Cook reductions \cite{pag06,alv93,val79b}. These problems were initially shown to be in $\totp$ by \Cref{sel-red} \cite{pag06}, which gives another way to prove that a function belongs to $\totp$. It suffices to prove that a function belongs to $\spe$ and it is self-reducible. In \cite{pag06}, it is shown that $\totp$ contains all self-reducible $\spe$-functions. 
	
	\begin{Theorem} [\cite{pag06}] \label{sel-red}
		$\totp$ is the closure under Karp reductions of the class of self-reducible $\spe$ functions. 
	\end{Theorem}
	
	The second condition in \Cref{pred-def}, which states that the existence of a witness with a given prefix can be determined in polynomial time, can be replaced by equivalent conditions:
	\begin{itemize}[\textbullet]
		\item There exists a polynomial-time algorithm that on input $(x, c) \in \Sigma^{*}$ outputs a witness for $x$ that is lexicographically closest to a given string $c$, or outputs that there is no witness for $x$.
		\item There exists a polynomial-time algorithm that on input $(x, a, b) \in \Sigma^{*}$ determines the existence of a witness for $x$ in a given interval $(a, b)$.
	\end{itemize}
	
	The above characterizations are similar to the characterization of $\totp$ in terms of interval-size functions defined on total $p$-orders with efficiently computable lexicographically nearest function (see \cite{bam16} for precise definitions).
	
	\begin{Theorem} [\cite{bam16}]
		$\totp = \mathsf{IF^{LN}_{t}}$. 
	\end{Theorem}
	
	\section{Closure properties of TotP} \label{sec-clo}
	
	In this section, we present closure properties of $\totp$ that complement the results from \cite{bak23} (\Cref{prop-cite}).
	
	\begin{Definition}
		For integers $a$ and $b$, the \textit{decrement} operation is defined as follows: 
		\[a \ominus b =
		\begin{cases}
			a - b, &\text{if $a \geqslant b$;}\\
			0, &\text{otherwise.}
		\end{cases}\] 
	\end{Definition}	
	
	\begin{Proposition} [\cite{bak23}]  \label[Proposition]{prop-cite}
		1. For any functions $f, g \in \totp$, the functions $f + g$, $f \cdot g$, and $f \ominus 1$ are in $\totp$.
		
		2. Let $f \in \totp$. If there exists a polynomial $p$ such that for any $x \in \Sigma^{*}$: $f(x) \leqslant p(|x|)$, then $f \in \fp$. 
	\end{Proposition}
	
	The following properties are not known to hold for $\sp$ and $\gapp$:
	
	\begin{Proposition} \label[Proposition]{prop-new}
		For any functions $f \in \totp$ and $g \in \fpl$, such that $g(x) \leqslant poly(|x|)$:
		\begin{enumerate}
			\item The languages $\{x \ | \ f(x) = g(x)\}$ and $\{x \ | \ f(x) > g(x)\}$ are in $\p$.
			\item The function $f(x) \ominus g(x)$ is in $\totp$. 
		\end{enumerate}
	\end{Proposition}
	
	The properties from \Cref{pow-bin} also hold for the classes $\sp$ \cite{bei92} and $\gapp$ \cite{fen94}. 
	
	\begin{Proposition} \label[Proposition]{pow-bin}
		For any functions $f \in \totp$ and $g \in \fpl$, such that $g(x) \leqslant poly(|x|)$:
		\begin{enumerate}
			\item The function $f(x)^{g(x)}$ is in $\totp$.
			\item The function $\binom{f(x)}{g(x)}$ is in $\totp$. 
		\end{enumerate}
	\end{Proposition}
	
	\begin{proof} 
		Let $f(x) = tot_{M}(x)$ for a NPTM $M$.
		\begin{enumerate}
			\item Let $N$ be the NPTM that behaves as follows on input $x \in \Sigma^{*}$:
			\begin{enumerate}[(a)]
				\item If $f(x) = 0$, $N$ halts.
				\item N takes one nondeterministic step and halts on the first path. On the second path, $N$ guesses $g(x)$ paths of $M(x)$ (with duplicates) that are lexicographically larger than the first path of $M(x)$.
			\end{enumerate} 	
			$tot_{N}(x) = f(x)^{g(x)}$.		
			\item Let $N$ be the NPTM that behaves as follows on input $x \in \Sigma^{*}$:
			\begin{enumerate}[(a)]
				\item If $f(x) = 0$, $N$ halts.
				\item If $M(x)$ has at least $g(x) + 1$ paths, $N$ takes one nondeterministic step and halts on the first path. On the second path, $N$ guesses a set of $g(x)$ distinct paths of $M(x)$ that are lexicographically larger than the first path of $M(x)$. 
			\end{enumerate}
			$tot_{N}(x) = \binom{f(x)}{g(x)}$. \qed
		\end{enumerate}
	\end{proof}
	
	The classes $\sp$ and $\gapp$ are closed under polynomial summation and polynomial product \cite{bei92,fen94}. The same is true for $\totp$ (\Cref{pol-sum}). 
	
	\begin{Proposition} \label[Proposition]{pol-sum}
		For any functions $f \in \totp$ and $g \in \fpl$, such that $g(x) \leqslant poly(|x|)$:
		\begin{enumerate}
			\item The function $s(x) = \sum_{0 \leqslant y \leqslant g(x)} f(x, y)$ is in $\totp$.
			\item The function $m(x) = \prod_{0 \leqslant y \leqslant g(x)} f(x, y)$ is in $\totp$. 
		\end{enumerate}
	\end{Proposition}
	
	\begin{proof}
		Let $f(x, y) = tot_{M}(x, y)$ for a NPTM $M$. 
		\begin{enumerate}
			\item Let $N$ be the NPTM that behaves as follows on input $x \in \Sigma^{*}$:
			\begin{enumerate}
				\item[] For $y$ in range $\{0, \ldots, g(x)\}$:
				\begin{enumerate}[leftmargin=1.5cm]
					\item[(a)] If $f(x, y) = 0$, $N$ proceeds to the next iteration of the loop.
					\item[(b)] If $f(x, y) > 0$, $N$ takes one nondeterministic step. On the first path, $N$ guesses one path of $M(x, y)$ that is lexicographically larger than the first path of $M(x, y)$.
					\item[(c)] If $y = g(x)$, $N$ halts on the second path. Otherwise, $N$ performs the next iteration of the loop on the second path.
				\end{enumerate}
			\end{enumerate}
			$tot_{N}(x) = \sum_{0 \leqslant y \leqslant g(x)} f(x, y)$. 	
			\vspace{0.8 mm}		
			\item Let $N$ be the NPTM that behaves as follows on input $x \in \Sigma^{*}$:
			\begin{enumerate}[leftmargin=1.5cm]
				\item[(a)] For $y$ in range $\{0, \ldots, g(x)\}$:
				\begin{enumerate}[leftmargin=1cm]
					\item[] If $f(x, y) = 0$, $N$ halts.
				\end{enumerate}
				\item[(b)] $N$ takes one nondeterministic step, halts on the first path, and proceeds on the second path.
				\item[(c)] For $y$ in range $\{0, \ldots, g(x)\}$:
				\begin{enumerate}[leftmargin=1cm]
					\item[] $N$ guesses one path of $M(x, y)$ that is lexicographically larger than the first path of $M(x, y)$.
				\end{enumerate}
			\end{enumerate}
			$tot_{N}(x) = \prod_{0 \leqslant y \leqslant g(x)} f(x, y)$. \qed
		\end{enumerate}
	\end{proof} 
	
	The classes $\sp$ and $\gapp$ are also closed under exponential summation \cite{bei92,fen94}. In \Cref{exp-sum} we prove that $\totp$ doesn't have this property unless $\p = \np$. 
	
	\begin{Proposition} \label[Proposition]{exp-sum}
		The following statements are equivalent:
		\begin{enumerate}
			\item For any functions $f \in \totp$ and $g \in \fpl$, such that $g(x) \leqslant poly(|x|)$, the function $s(x) = \sum_{|y| \leqslant g(x)} f(x, y)$ is in $\totp$.
			\item $\p = \np$. 
		\end{enumerate}
	\end{Proposition}
	
	\begin{proof} 
		We will use the equivalent $\totp = \sp$ instead of $\p = \np$. If $\totp = \sp$, then the first statement holds, because $\sp$ has this property \cite{bei92}. Let $h$ be a function in $\sp$. There exist an $\fp$-function $V$$:\Sigma^{*} \times \Sigma^{*} \rightarrow \{0, 1\}$ and a polynomial $p$ such that for any $x \in \Sigma^{*}$: $h(x) = |\{y \ | \ |y| \leqslant p(|x|) \wedge V(x, y) = 1\}|$. By assumption, the function $h(x) = \sum_{|y| \leqslant p(|x|)} V(x, y)$ is in $\totp$. \qed
	\end{proof}

	\section{Characterization of GapP with TotP functions} \label{sec-gap}
	
	In this section, we give a more precise characterization of $\gapp$ with $\totp$-functions, than $\gapp = \totp - \fp$ \cite{bak23}, given in \Cref{rel-fun}. In the second part of \Cref{gap-lem} the class $\totp$ can be replaced by the class $\spanl$ \cite{alv93} (the proof is given in \cite{hem95}). $\spanl$ is a subclass of $\totp$, so the proof of the second part of \Cref{gap-lem} also follows from \cite{hem95}.  We give a simpler proof of this statement.
	
	\begin{Lemma} \label[Lemma]{gap-lem}
		For any NPTM $M$ and any polynomial $p$ such that for any $x \in \Sigma^{*}$ $p(|x|)$ is greater than the running time of $M(x)$:
		\begin{enumerate}
			\item \textnormal{\cite{kia01}} There exists a function $g_{1} \in \totp$ such that for any $x \in \Sigma^{*}$: $acc_{M}(x) = g_{1}(x) - 2^{p(|x|)}$.			
			\item There exists a function $g_{2} \in \totp$ such that for any $x \in \Sigma^{*}$: $acc_{M}(x) = 2^{p(|x|)} - g_{2}(x)$. 
		\end{enumerate}
	\end{Lemma}
	
	\begin{proof}
		\begin{enumerate}
			\item We repeat the proof from \cite{kia01}. There exists a NPTM $M'$ such that $acc_{M}(x) = acc_{M'}(x)$ and $Total_{M'}(x) = 2^{p(|x|)}$. Let $N$ be the NPTM that simulates $M'$, but instead of going into an accepting state, generates two accepting paths. \\ $acc_{M}(x) = Total_{N}(x) - Total_{M'}(x) = (tot_{N}(x) + 1) - 2^{p(|x|)}$.
			\item There exists a NPTM $M'$ such that $acc_{M}(x) = acc_{M'}(x)$ and $Total_{M'}(x) = 2^{p(|x|) - 1}$. Let $N$ be the NPTM that simulates $M'$, but instead of going into a rejecting state, generates two rejecting paths. \\
			$acc_{M}(x) = 2 \cdot Total_{M'}(x) - Total_{N}(x) = 2^{p(|x|)} - (tot_{N}(x) + 1)$. \qed
		\end{enumerate}
	\end{proof}
	
	\begin{Proposition}
		For any function $f \in \gapp$ there exists a function $g \in \totp$ and a polynomial $p$ such that for any $x \in \Sigma^{*}$: $f(x) = g(x) - 2^{p(|x|)}$. 
	\end{Proposition}
	
	\begin{proof}
		Let $f(x) = acc_{M}(x) - acc_{N}(x)$ for NPTMs $M$ and $N$. By \Cref{gap-lem}, there exist functions $g_{1}, g_{2} \in \totp$ such that $acc_{M}(x) = g_{1}(x) - 2^{p(|x|)}$ and $acc_{N}(x) = 2^{p(|x|)} - g_{2}(x)$, where $p$ is a polynomial that is greater than the running time of $M$ and $N$. \\
		$f(x) = (g_{1}(x) - 2^{p(|x|)}) - (2^{p(|x|)} - g_{2}(x)) = (g_{1}(x) + g_{2}(x)) - 2^{p(|x|) + 1}$. \qed
	\end{proof}
	
	In \cite{for97}, it is shown that for every language $L \in \cp$ there exists a function $f \in \gapp$ such that for any $x \in \Sigma^{*}$:
	\begin{align*}
		x \in L \Leftrightarrow f(x) = 0, \\
		x \notin L \Leftrightarrow f(x) < 0.
	\end{align*}
	
	The following characterization of $\cp$ we use in the next section:
	
	\begin{Lemma} \label[Lemma]{cp-def}
		For any language $L \in \cp$ there exist a function $g \in \totp$ and a polynomial $p$ such that for any $x \in \Sigma^{*}$:
		\begin{align*}
			x \in L \Leftrightarrow g(x) = 2^{p(|x|)}, \\
			x \notin L \Leftrightarrow g(x) < 2^{p(|x|)}. 
		\end{align*}
	\end{Lemma}	
	
	\section{Closure under composition with FP} \label{sec-comp}
	
	In this section, we prove that the closure of $\totp$ under left composition with $\fpl$ is equivalent to $\totp = \fpl$ and $\pp = \p$. In the next theorem, we will also consider $\fpl$-functions of several arguments. A function $h$ is in the class $\fpl \circ (\totp \times ... \times \totp)$ if there exist functions $f \in \fpl$ and $g_{1}, \ldots, g_{k} \in \totp$ such that for any $x \in \Sigma^{*}$: $h(x) = f(g_{1}(x), \ldots, g_{k}(x))$. 
		
	\begin{Theorem} \label{comp-totp}
		The following statements are equivalent:
		\begin{enumerate}
			\item $\fpl \circ \totp \subseteq \totp$.
			\item $\fpl \circ (\totp \times \ldots \times \totp) \subseteq \totp$.
			\item $\totp = \fpl$.
			\item $\pp = \p$. 
		\end{enumerate}
	\end{Theorem}
	
	\begin{proof}
		$(2 \Rightarrow 1)$: (1) is equivalent to (2), because the classes $\fpl \circ \totp$ and  $\fpl \circ (\totp \times \ldots \times \totp)$ are equal \cite{iva25}.
				
		$(1 \Rightarrow 4)$: Let $L = \{x \ | \ f(x) > g(x)\}$, where $f \in \totp$, $g \in \fp$.
		The function 
		\begin{align*}
			h(x) =
			\begin{cases}
				1, &\text{$f(x) > g(x)$;}\\
				0, &\text{$\textnormal{otherwise}$.}
			\end{cases}
		\end{align*}
		is in $\fpl^{\totp[1]}$, which is equal to the class $\fpl \circ \totp$ \cite{iva25}. By assumption, $h \in \totp$. By \Cref{prop-cite}, $h \in \fp$ and $L \in \p$. 
		
		$(4 \Rightarrow 3)$: $\totp = \fpl$ is equivalent to $\sp = \fpl$, which is well known to be equivalent to $\pp = \p$.
		
		$(3 \Rightarrow 2)$: Follows from $\fpl \circ (\fpl \times \ldots \times \fpl) \subseteq \fpl$. \qed
	\end{proof}
	
	For the class $\sp$, there exists $\fpl$-functions of several arguments such that if $\sp$ is closed under composition with them, then it is closed under composition with $\fpl$ \cite{ogi93}. In the next theorem, we show that these functions have the same property when applied to the class $\totp$. We repeat the proofs from \cite{ogi93} that are slightly modified for the case of $\totp$.
	
	\begin{Definition}
		\begin{enumerate}
			\item $span(a_{1}, \ldots, a_{k})$ is the number of distinct elements in a set $\{a_{1}, \ldots, a_{k}\}$.			
			\item $plu(a_{1}, \ldots, a_{k})$ is the set of most frequently occurring elements in a set $\{a_{1}, \ldots, a_{k}\}$. 
		\end{enumerate}
	\end{Definition}
	
	\begin{Theorem}
		The following statements are equivalent:
		\begin{enumerate}
			\item $\fpl \circ (\totp \times \ldots \times \totp) \subseteq \totp$.
			\item $\cp = \p$.
			\item For any functions $f\in \totp$, $g \in \fp$, the function $f(x) \ominus g(x)$ is in $\totp$.
			\item For any functions $f\in \totp$, $g \in \fpl$, such that $g(x) \neq 0$, the function 
			$\lfloor f(x) / g(x) \rfloor$ is in $\totp$.
			\item For any functions $f_{1}, \ldots, f_{k} \in \totp$ the function $g(x) = span(f_{1}(x), \ldots, f_{k}(x))$ is in $\totp$.
			\item For any functions $f_{1}, \ldots, f_{k} \in \totp$ there exists a function $g \in \totp$ such that for any $x \in \Sigma^{*}$: $g(x) \in plu(f_{1}(x), \ldots, f_{k}(x))$. 
		\end{enumerate}
	\end{Theorem}
	
	\begin{proof}
		(1) and (2) are equivalent by \Cref{comp-totp} (instead of $\pp = \p$ we use the equivalent $\cp = \p$ \cite{ogi93}). (1) implies (3), (4), (5), and (6) because these functions are polynomial time computable (for (6) it suffices to consider the function $min(plu(a_{1}, \ldots, a_{k}))$). 
		
		Let $L \in \cp$. By \Cref{cp-def}, there exist a function $f \in \totp$ and a polynomial $p$ such that for any $x \in \Sigma^{*}$:
		\begin{align*}
			x \in L \Leftrightarrow f(x) = 2^{p(|x|)}, \\
			x \notin L \Leftrightarrow f(x) < 2^{p(|x|)}. 
		\end{align*}
		
		$(3 \Rightarrow 2)$: Let $g(x) = f(x) \ominus (2^{p(|x|)} - 1)$. By assumption, $g \in \totp$, and for any $x \in \Sigma^{*}$:
		\begin{align*}
			x \in L \Leftrightarrow g(x) = 1, \\
			x \notin L \Leftrightarrow g(x) = 0.
		\end{align*}
		By \Cref{prop-cite}, $g \in \fp$ and $L \in \p$. 
		
		$(4 \Rightarrow 2)$: Let $g(x) = \lfloor f(x) / 2^{p(|x|)} \rfloor$. By assumption, $g \in \totp$, and for any $x \in \Sigma^{*}$:
		\begin{align*}
			x \in L \Leftrightarrow g(x) = 1, \\
			x \notin L \Leftrightarrow g(x) = 0.
		\end{align*}
		By \Cref{prop-cite}, $g \in \fp$ and $L \in \p$. 
		
		$(5 \Rightarrow 2)$: Let $g(x) = span(f(x), 2^{p(|x|)})$. By assumption, $g \in \totp$, and for any $x \in \Sigma^{*}$:
		\begin{align*}
			x \in L \Leftrightarrow g(x) = 1, \\
			x \notin L \Leftrightarrow g(x) = 2.
		\end{align*} 
		By \Cref{prop-cite}, $g \in \fp$ and $L \in \p$. 
		
		$(6 \Rightarrow 2)$: By assumption, there exists a function $g \in \totp$ such that for any $x \in \Sigma^{*}$: $g(x) \in plu(0, 0, 2f(x), f(x) + 2^{p(|x|)}, 2^{p(|x|) + 1})$,
		\begin{align*}
			x &\in L \Leftrightarrow g(x) = 2^{p(|x|) + 1}, \\
			x &\notin L \Leftrightarrow g(x) = 0.
		\end{align*}
		By \Cref{prop-new}, $L \in \p$. \qed
	\end{proof}

	
	
	
	
	%
	%
	%
	\bibliographystyle{splncs04}
	\bibliography{bibliography_totp}

\end{document}